# Guardians of the Web: The Evolution and Future of Website Information Security

Md Saiful Islam[1,2], Li Xiangdong[1,2, *]

1 School of Computer Science, Zhongyuan University of Technology, Zhengzhou, China

2 Henan Key Laboratory on Public Opinion Intelligent Analysis, Zhengzhou, China

*Correspondence Author: lixiangdong@zut.edu.cn

## Abstract

Website information security has become a critical concern in the digital age. This article explores the evolution of website information security, examining its historical development, current practices, and future directions. The early beginnings from the 1960s to the 1980s laid the groundwork for modern cybersecurity, with the development of ARPANET, TCP/IP, public-key cryptography, and the first antivirus programs. The 1990s marked a transformative era, driven by the commercialization of the Internet and the emergence of web-based services. As the Internet grew, so did the range and sophistication of cyber threats, leading to advancements in security technologies such as the Secure Sockets Layer (SSL) protocol, password protection, and firewalls. Current practices in website information security involve a multi-layered approach, including encryption, secure coding practices, regular security audits, and user education. The future of website information security is expected to be shaped by emerging technologies such as artificial intelligence, blockchain, and quantum computing, as well as the increasing importance of international cooperation and standardization efforts. As cyber threats continue to evolve, ongoing research and innovation in website information security will be essential to protect sensitive information and maintain trust in the digital world.

## 1. Introduction

In the digital age, website information security has become a critical concern for businesses, governments, and individuals alike. As the Internet continues to grow and integrate into every aspect of our lives, protecting sensitive information from cyber threats is paramount. This document, "Guardians of the Web: The Evolution and Future of Website Information Security," explores the historical development, current practices, and future directions of website information security[1].

The journey begins with the early days of networked computing from the 1960s to the 1980s. During this period, foundational technologies like ARPANET, TCP/IP, and public-key cryptography were developed, setting the stage for modern cybersecurity measures. The creation of the first antivirus programs and the establishment of basic security principles marked significant milestones in the fight against emerging cyber threats.

The 1990s marked a transformative era driven by the commercialization of the Internet and the rapid growth of web-based services. This decade saw the introduction of critical security technologies such as the Secure Sockets Layer (SSL) protocol, which provided a means to encrypt data transmitted between web servers and browsers[2], [3]. As the Internet evolved, so did the range and sophistication of cyber threats, necessitating advancements in security measures like password protection and firewalls.

Moving into the 2000s, the rise of complex malware and distributed denial-of-service (DDoS) attacks prompted further advancements in cybersecurity. This period witnessed the development of sophisticated firewalls, intrusion detection systems (IDS), and intrusion prevention systems (IPS). The adoption of encryption technologies like Transport Layer Security (TLS) became widespread, ensuring the confidentiality and integrity of data transmitted over the Internet.

In the contemporary era, cybersecurity practices have evolved dramatically to address the increasing sophistication and frequency of cyber threats. The adoption of Zero Trust Architecture, which emphasizes continuous verification of user identities and stringent access controls, represents a fundamental shift in security paradigms[4]. Artificial Intelligence (AI) and Machine Learning (ML) have become transformative technologies in cybersecurity, enhancing threat detection and response capabilities through real-time analysis and predictive analytics.

Cloud security has emerged as a critical focus due to the widespread adoption of cloud services. Tools like Cloud Security Posture Management (CSPM) and Cloud Access Security Brokers (CASBs) play essential roles in monitoring and securing cloud environments. The integration of security into the DevOps workflow, known as DevSecOps, ensures that security is considered at every stage of the software development lifecycle[5].

Data privacy and protection have been strengthened through robust encryption practices, Data Loss Prevention (DLP) tools, and adherence to stringent regulations such as the General Data Protection

Regulation (GDPR) and the California Consumer Privacy Act (CCPA). These measures collectively enhance the ability of organizations to protect their digital assets and ensure the integrity, confidentiality, and availability of their information.

Looking to the future, emerging technologies like quantum computing, blockchain, and advanced encryption techniques promise to revolutionize cybersecurity[6]. Quantum computing, while posing a risk to traditional encryption methods, drives the development of quantum-resistant algorithms and advanced cryptographic techniques like Quantum Key Distribution (QKD). Blockchain technology offers robust security features for identity management, supply chain integrity, and secure transactions.

Artificial Intelligence (AI) and Machine Learning (ML) will continue to transform cybersecurity by enabling more sophisticated and adaptive defense mechanisms. These technologies can analyze massive datasets to identify patterns and anomalies that may indicate cyber threats, often in real-time. The rapid expansion of the Internet of Things (IoT) introduces new security challenges[7], [8], requiring robust security protocols and network segmentation to protect critical systems.

In addition to technological advancements, the future of cybersecurity will increasingly depend on comprehensive regulatory frameworks and international cooperation. Standardized security practices and policies across different countries and industries are essential to address the global nature of cyber threats. International collaboration through entities like the United Nations and the Council of Europe's Convention on Cybercrime will play a crucial role in harmonizing cybersecurity laws and coordinating responses to cross-border cyber threats.

## 2. Historical Development

### 2.1 Early Beginnings (1960s-1980s)

The early beginnings of website information security, from the 1960s to the 1980s, were instrumental in laying the groundwork for the advanced cybersecurity measures we rely on today. During the 1960s, the concept of networked computing began to materialize with the development of ARPANET, initiated by the U.S[9]. Department of Defense's Advanced Research Projects Agency (ARPA). ARPANET aimed to connect various academic and research institutions, allowing them to share information and resources efficiently. In these early days, security concerns primarily revolved around physical access to computers and ensuring that only authorized personnel could use these powerful machines[10], [11].

As ARPANET expanded in the 1970s, connecting more institutions, the network's vulnerabilities began to surface[12]. One of the first significant security incidents was the creation of the "Creeper" program by Bob Thomas in 1971, which could move across the network and leave a trace message. Although Creeper was not malicious, it demonstrated the potential for network-based threats. In response, Ray Tomlinson developed "Reaper," the first known antivirus program, designed to locate and delete Creeper[13]. This period also saw the establishment of foundational security principles and technologies, such as the development of the Transmission Control Protocol (TCP) and Internet Protocol (IP) by Vint Cerf and Bob Kahn in 1974. TCP/IP provided a standardized method for data transmission across interconnected networks, forming the basis of the modern Internet[14], [15].

The 1970s also marked significant advancements in cryptography, crucial for securing communications. Whitfield Diffie and Martin Hellman's 1976 paper introduced the concept of public-key cryptography, which allowed secure communication between parties without a shared secret key. Their Diffie-Hellman key exchange protocol laid the foundation for many encryption protocols used today[16]. Additionally, the MULTICS (Multiplexed Information and Computing Service) operating system, developed by MIT, Bell Labs, and General Electric, introduced advanced security features like hierarchical file systems and access control lists (ACLs), influencing the design of later secure operating systems.

The 1980s saw the proliferation of personal computers and the expansion of computer networks, increasing the exposure to cyber threats. The rise of personal computing brought about a new era of interconnectedness, but it also made systems more vulnerable to attacks. One of the most notable security incidents of the decade was the 1983 hacking of U.S. military networks by a group of German hackers, which brought international attention to the issue of network security[17]. This period also witnessed the development of the Data Encryption Standard (DES) by the National Institute of Standards and Technology (NIST) in 1977, which became a widely adopted method for encrypting electronic data, despite its later vulnerabilities to brute-force attacks.

The 1980s also saw the birth of the computer security industry. Companies like McAfee and Norton emerged, offering antivirus software to combat the growing threat of computer viruses. These early antivirus programs relied on signature-based detection methods, where known virus signatures were matched against files on a user's computer. This method was effective for known threats but struggled to keep pace with the rapidly evolving malware landscape.

In response to the growing threat of cyber-crime, governments began to develop policies and legislation to address these new challenges[18]. The United States enacted the Computer Fraud and Abuse Act (CFAA) in 1986, making unauthorized access to computer systems a federal crime. This legislation provided a legal framework for prosecuting cybercriminals and highlighted the increasing importance of cybersecurity.

The period from the 1960s to the 1980s was marked by significant advancements in networked computing and the initial development of security measures. The creation of ARPANET, the introduction of TCP/IP, the development of public-key cryptography, and the establishment of the first antivirus programs all contributed to the foundation of modern cybersecurity. These early efforts laid the groundwork for the sophisticated security practices and technologies that would emerge in the following decades, as the world became increasingly dependent on digital information and communication.

## 2.2 Emergence of the Internet and Basic Security (1990s)

The 1990s marked a transformative era in the development of website information security, driven by the commercialization of the Internet and the rapid proliferation of web-based services. This decade saw the Internet evolve from a primarily academic and governmental network into a global commercial platform accessible to the general public. The introduction of the World Wide Web by Tim Berners-Lee in 1991 fundamentally changed how information was shared and accessed, leading to a dramatic increase in online activity and the establishment of numerous websites and e-commerce platforms[19].

As the Internet grew, so did the range and sophistication of cyber threats. The early 1990s witnessed a surge in malicious activities, including the spread of viruses, worms, and the first large-scale distributed denial-of-service (DDoS) attacks. One of the most notable incidents was the "ILOVEYOU" virus in 2000, which caused widespread damage by spreading through email systems and affecting millions of computers worldwide[20], [21]. This highlighted the vulnerabilities in online communication systems and the need for robust security measures.

To address these emerging threats, significant advancements in security technologies were made. One of the most critical developments was the introduction of the Secure Sockets Layer (SSL) protocol by Netscape Communications in 1994. SSL provided a means to encrypt data transmitted between web servers and browsers, ensuring the confidentiality and integrity of online transactions. This technology was crucial for the growth of e-commerce, as it enabled secure online shopping and banking by protecting sensitive information such as credit card numbers and personal details[22].

In addition to SSL, basic security measures like password protection and firewalls became standard practices. Password protection was implemented to restrict access to sensitive information, although the effectiveness of this measure was often undermined by weak password policies and user practices. Firewalls, which serve as barriers between internal networks and external threats, evolved from simple packet-filtering devices to more sophisticated systems capable of stateful inspection, which tracks the state of active connections and makes decisions based on the context of the traffic.

The antivirus software industry also took shape during this period, with companies like McAfee and Symantec developing tools to detect, quarantine, and remove malicious software. These antivirus programs relied on signature-based detection, where known malware signatures were matched against files on users' computers[23]. Although effective against known threats, this method struggled to keep up with the rapid emergence of new malware variants.

Regulatory and policy frameworks began to emerge in response to the growing importance of cybersecurity. In the United States, the Health Insurance Portability and Accountability Act (HIPAA) was enacted in 1996, establishing national standards for the protection of health information. HIPAA mandated that healthcare organizations implement security measures to safeguard patient data, marking one of the first significant regulatory efforts to address information security.

The establishment of the Open Web Application Security Project (OWASP) in 2001, although slightly outside the 1990s timeframe, was rooted in the security challenges of the previous decade. OWASP provided guidelines and tools to help developers identify and mitigate common vulnerabilities in web applications, such as SQL injection and cross-site scripting (XSS). These efforts underscored the importance of securing web applications, which had become increasingly targeted by cybercriminals[24].

The 1990s laid the foundational groundwork for modern cybersecurity practices. The rapid expansion of the Internet and the rise of cyber threats prompted significant advancements in encryption technologies, the development of antivirus software, and the implementation of basic security measures like firewalls and password protection. Regulatory frameworks began to take shape, emphasizing the importance of protecting sensitive information. These developments collectively underscored the critical need to secure digital information and communications in an increasingly interconnected world, setting the stage for the sophisticated cybersecurity measures that would follow in the subsequent decades.

## 2.3 Rise of Cyber Threats and Advanced Security Measures (2000s)

The 2000s witnessed a dramatic escalation in the variety and sophistication of cyber threats, compelling the cybersecurity industry to develop advanced security measures in response. The decade began with the "ILOVEYOU" virus in 2000, which caused widespread havoc by exploiting email systems to propagate itself, highlighting the urgent need for better email security practices. This period saw the emergence of more complex malware, such as the "Code Red" worm in 2001 and the "Slammer" worm in 2003, which exploited vulnerabilities in web servers and database systems, respectively, causing significant disruptions and financial losses[25], [26]. The increasing prevalence of Distributed Denial of Service (DDoS) attacks, notably the 2000 attack on Yahoo, underscored the vulnerability of high-profile websites to coordinated assault by botnets.

In response to these growing threats, the cybersecurity landscape evolved rapidly. Firewalls advanced from simple packet-filtering devices to sophisticated systems capable of deep packet inspection and application-level filtering. Intrusion Detection Systems (IDS) and Intrusion Prevention Systems (IPS) became integral components of network security, capable of monitoring network traffic in real-time to detect and mitigate suspicious activities[27]. The adoption of encryption technologies like Transport Layer Security (TLS), which succeeded Secure Sockets Layer (SSL), became widespread, ensuring the confidentiality and integrity of data transmitted over the Internet.

The antivirus industry also evolved, with traditional signature-based detection being supplemented by heuristic and behavior-based detection techniques, allowing for the identification of previously unknown threats. The rise of Security Information and Event Management (SIEM) systems enabled organizations to collect, analyze, and correlate security data from various sources, providing a comprehensive view of their security posture and facilitating faster incident response.

Regulatory frameworks were strengthened during this decade to address the growing cyber threat landscape. The enactment of laws such as the Sarbanes-Oxley Act (SOX) in 2002 and the Payment Card Industry Data Security Standard (PCI DSS) in 2004 established stringent security requirements for financial and payment card information[28], driving organizations to adopt more rigorous security measures. International efforts to harmonize cybersecurity laws also gained momentum, with initiatives like the Council of Europe's Convention on Cybercrime promoting global cooperation in combating cyber threats[29].

Furthermore, the rise of cybercrime as a service, where malicious actors offered hacking tools and services for hire, exacerbated the threat landscape. This period saw the proliferation of phishing and spear-phishing attacks, which became increasingly sophisticated in targeting individuals and organizations to steal sensitive information.

The 2000s were marked by significant advancements in cybersecurity technology and practices, driven by the escalating complexity and frequency of cyber threats. The development of advanced firewalls, IDS/IPS systems, and SIEM solutions, along with stronger regulatory frameworks, laid the groundwork for modern cybersecurity defenses. These efforts collectively enhanced the ability of organizations to protect their digital assets in an increasingly hostile cyber environment, setting the stage for further advancements in the following decades.

## 3. Contemporary Security Practices

### 3.1 Modern Era of Cybersecurity (2010s-Present)

From the 2010s to the present, cybersecurity practices have evolved dramatically, driven by the increasing sophistication and frequency of cyber threats. This era has seen the widespread adoption of Zero Trust Architecture, which fundamentally shifts security paradigms by operating on the principle of "never trust, always verify." Unlike traditional perimeter-based security models, Zero Trust requires continuous verification of user identities and stringent access controls regardless of the user's location or device, significantly reducing the risk of unauthorized access[30].

Artificial Intelligence (AI) and Machine Learning (ML) have become transformative technologies in cybersecurity, enhancing threat detection and response capabilities. These technologies analyze vast amounts of data to identify patterns and anomalies that indicate potential cyber threats. AI-driven solutions can detect zero-day exploits and sophisticated attacks by learning from historical data and predicting potential vulnerabilities. Machine learning algorithms continuously improve their detection methods based on new data inputs, enabling real-time threat detection and automated responses that can isolate compromised systems or block malicious activities promptly.

Cloud security has emerged as a critical focus due to the widespread adoption of cloud services. Cloud Security Posture Management (CSPM) tools have become essential for continuously monitoring cloud infrastructure for misconfigurations and compliance violations[31]. These tools provide automated remediation suggestions, ensuring a secure cloud environment. Cloud Access Security Brokers (CASBs) act as intermediaries between users and cloud service providers, enforcing security policies, ensuring compliance, and providing visibility into cloud application usage. Secure Access Service Edge (SASE) frameworks integrate networking and security functions into a unified, cloud-based service, offering secure, optimized access to cloud applications and protecting remote workforces[32].

DevSecOps represents the integration of security into the DevOps workflow, promoting collaboration between development, security, and operations teams. This approach ensures that security is considered at every stage of the software development lifecycle. Key practices include automated security testing within continuous integration/continuous deployment (CI/CD) pipelines, continuous monitoring for security threats, and applying security controls to Infrastructure as Code (IaC)[19]. These measures help identify and fix vulnerabilities early in the development process, improving the overall security and reliability of applications.

Endpoint security has advanced significantly with the adoption of Endpoint Detection and Response (EDR) solutions and Next-Generation Antivirus (NGAV) tools. EDR solutions provide continuous monitoring and analysis of endpoint activities, enabling security teams to detect, investigate, and respond to threats in real-time. NGAV solutions go beyond traditional signature-based detection by using machine learning and behavioral analysis to identify and block advanced malware, including zero-day exploits. Unified Endpoint Management (UEM) platforms enable centralized management and security of all endpoints, including desktops, laptops, mobile devices, and IoT devices, ensuring consistent security policies across the organization[33].

Data privacy and protection have been strengthened through robust encryption practices, Data Loss Prevention (DLP) tools, and adherence to stringent regulations such as the General Data Protection Regulation (GDPR) and the California Consumer Privacy Act (CCPA). GDPR and CCPA enforce strict data protection standards and grant consumers rights over their personal data, compelling organizations to implement comprehensive data protection measures. These regulations require organizations to ensure the confidentiality, integrity, and availability of personal data, driving the adoption of robust security practices.

Regulatory frameworks like the National Institute of Standards and Technology (NIST) Cybersecurity Framework provide guidelines and best practices for managing cybersecurity risks. The NIST framework's core functions—Identify, Protect, Detect, Respond, and Recover—help organizations build comprehensive cybersecurity programs. By adopting these frameworks, organizations can systematically enhance their cybersecurity posture, ensuring they are prepared to prevent, detect, and respond to cyber threats effectively.

The use of AI-driven threat detection, automation, and orchestration has streamlined security operations, improving efficiency and response times. Security Orchestration, Automation, and Response (SOAR) platforms integrate with various security tools, enabling automated workflows for incident detection, investigation, and remediation[34]. These platforms reduce the manual effort required for security operations, allowing security teams to focus on more complex and strategic tasks.

User awareness and training programs have also become essential components of contemporary security practices. Organizations implement regular training sessions, phishing simulations, and awareness campaigns to educate employees about cybersecurity risks and best practices. Role-based training ensures that employees understand the specific security threats relevant to their roles and responsibilities, fostering a security-conscious culture.

The modern era of cybersecurity is characterized by the integration of advanced technologies, comprehensive regulatory frameworks, and proactive security measures. These practices collectively enhance the ability of organizations to protect their digital assets and ensure the integrity, confidentiality, and availability of their information in an increasingly complex and interconnected cyber landscape.

## 3.2 Current Trends

The current trends in contemporary security practices showcase a multi-faceted approach to tackling the increasing complexity of cyber threats. Organizations are increasingly adopting Zero Trust Architecture, a security model based on the principle of "never trust, always verify." This approach assumes that threats can come from both internal and external sources, necessitating continuous verification of user identities and strict access controls, regardless of the user's location or device. This model typically involves micro-segmentation of networks, robust identity and access management (IAM) systems with multi-factor authentication (MFA), continuous monitoring for anomalies, and enforcing least privilege access policies[35].

Artificial intelligence (AI) and machine learning (ML) have become transformative tools in cybersecurity. These technologies analyze vast amounts of data to detect patterns and anomalies that traditional methods might miss. AI-driven threat detection systems can identify zero-day exploits and sophisticated attack techniques by learning from historical data and predicting potential vulnerabilities. Machine learning algorithms enhance these capabilities by continuously improving their detection methods based on new data inputs, enabling real-time threat detection and automated responses that can isolate compromised systems or block malicious activities promptly.

Cloud security has also become a critical focus as more organizations move their operations to cloud environments. Cloud Security Posture Management (CSPM) solutions continuously monitor cloud infrastructure for misconfigurations and compliance violations, providing automated remediation suggestions[36]. Cloud Access Security Brokers (CASBs) serve as intermediaries between users and cloud service providers, enforcing security policies, ensuring compliance, and providing visibility into cloud application usage. Secure Access Service Edge (SASE) frameworks integrate networking and security functions into a unified, cloud-based service, offering secure, optimized access to cloud applications and protecting remote workforces.

DevSecOps represents the integration of security into the DevOps workflow, promoting collaboration between development, security, and operations teams. This approach ensures that security is considered at every stage of the software development lifecycle. Key practices include automated security testing within continuous integration/continuous deployment (CI/CD) pipelines, continuous monitoring for security threats, and applying security controls to Infrastructure as Code (IaC). These measures help identify and fix vulnerabilities early in the development process, improving the overall security and reliability of applications[37].

Endpoint security has advanced with the adoption of Endpoint Detection and Response (EDR) solutions and Next-Generation Antivirus (NGAV) tools. EDR solutions provide continuous monitoring and analysis of endpoint activities, enabling security teams to detect, investigate, and respond to threats in real-time. NGAV solutions go beyond traditional signature-based detection by using machine learning and behavioral analysis to identify and block advanced malware, including zero-day exploits.

Data privacy and protection are increasingly emphasized through robust encryption practices, Data Loss Prevention (DLP) tools, and adherence to privacy-centric regulations such as the General Data Protection Regulation (GDPR) and the California Consumer Privacy Act (CCPA). Encryption ensures the confidentiality and integrity of data at rest and in transit, while DLP solutions prevent unauthorized data transfers and protect sensitive information[38]. Compliance with regulations like GDPR and CCPA necessitates stringent data protection measures and provides consumers with rights over their personal data.

Regulatory frameworks like the NIST Cybersecurity Framework offer guidelines and best practices for managing cybersecurity risks. The NIST framework's core functions—Identify, Protect, Detect, Respond, and Recover—help organizations build comprehensive cybersecurity programs. By adopting these frameworks, organizations can systematically enhance their cybersecurity posture, ensuring they are prepared to prevent, detect, and respond to cyber threats effectively.

These contemporary security practices and trends reflect the ongoing evolution of cybersecurity in response to the ever-changing threat landscape. By integrating advanced technologies, regulatory compliance, and proactive security measures, organizations can better protect their digital assets and ensure the integrity, confidentiality, and availability of their information.

# 4 Future Directions

## 4.1 Cybersecurity

The future of cybersecurity is being profoundly reshaped by several emerging technologies that are still in their nascent stages but show immense potential for revolutionizing how digital assets are protected. Quantum computing, although posing a significant risk to traditional encryption methods due to its unprecedented processing power, is simultaneously driving the development of quantum-resistant algorithms and advanced cryptographic techniques like Quantum Key Distribution (QKD)[39], which offers theoretically unbreakable encryption by leveraging the principles of quantum mechanics.

Blockchain technology, initially popularized by cryptocurrencies, is being adapted for a broad range of applications beyond financial transactions. Its decentralized, immutable ledger system provides robust security features ideal for enhancing identity management systems, ensuring the integrity and transparency of supply chains, and securing critical transactions through smart contracts, which automate and enforce agreements without human intervention.

Artificial Intelligence (AI) and Machine Learning (ML) are transforming cybersecurity by enabling more sophisticated and adaptive defense mechanisms[40]. These technologies can analyze massive datasets to identify patterns and anomalies that may indicate cyber threats, often in real-time. AI and ML are not only improving threat detection but also enabling predictive analytics that foresee potential vulnerabilities, allowing organizations to take proactive measures. Automated response capabilities provided by AI can isolate compromised systems and block malicious activities swiftly, reducing the time it takes to mitigate threats[41].

The rapid expansion of the Internet of Things (IoT) introduces new security challenges, as each connected device represents a potential entry point for cyber-attacks. Securing IoT devices requires embedding robust security protocols directly into the devices, ensuring secure communication channels, and employing network segmentation to protect critical systems. Advances in edge computing, which processes data closer to its source, are enhancing IoT security by reducing latency and enabling quicker threat detection and response.

Biometric authentication is emerging as a highly secure alternative to traditional password-based systems. By using unique biological traits such as fingerprints, facial recognition, or iris scans, biometric systems offer a higher level of security that is difficult to replicate or steal. When combined with multi-factor authentication (MFA), biometric systems significantly strengthen security by ensuring that access is granted only to authorized individuals.

Advanced encryption techniques are also evolving to address future security challenges. Post-quantum cryptography focuses on developing encryption methods that are resistant to the capabilities of quantum computers, ensuring that data remains secure in a post-quantum world. Homomorphic encryption, which allows computations to be performed on encrypted data without decrypting it, promises to enhance data security[42], particularly in cloud computing environments, by enabling secure data processing.

In addition to technological advancements, the future of cybersecurity will increasingly depend on comprehensive regulatory frameworks and international cooperation. As cyber threats become more sophisticated and global, standardized security practices and policies across different countries and industries are essential. Regulations like the General Data Protection Regulation (GDPR) and the California Consumer Privacy Act (CCPA) set high standards for data protection and have driven organizations worldwide to adopt robust security measures. International collaboration through entities like the United Nations and the Council of Europe's Convention on Cybercrime is crucial for harmonizing cybersecurity laws, facilitating information sharing, and coordinating responses to cross-border cyber threats.

These emerging technologies and evolving regulatory frameworks collectively promise to provide more robust, adaptive, and proactive security measures. They ensure that organizations can better protect their digital assets against an ever-evolving threat landscape, maintaining the integrity, confidentiality, and availability of their information in the face of increasingly sophisticated cyber threats.

## 4.2 Proactive Threat Management

The future of proactive threat management is being revolutionized by a range of advanced technologies and innovative strategies that aim to identify and mitigate cyber threats before they can inflict damage. Artificial Intelligence (AI) and Machine Learning (ML) play a pivotal role in this evolution, providing the capabilities for predictive analytics that forecast potential threats and automated threat hunting that continuously scans for and neutralizes suspicious activities. These technologies enable a shift from reactive to proactive security postures, identifying patterns and anomalies in vast datasets that could indicate looming threats.

Real-time threat intelligence has become crucial, with platforms that gather and analyze data from multiple sources, including global threat databases, dark web forums, and hacker communities. This real-time analysis allows organizations to stay ahead of emerging threats and dynamically adjust their defenses[43]. Behavioral analytics and User Behavior Analytics (UBA) are integral to detecting insider threats and advanced persistent threats (APTs) by establishing baselines of normal behavior and identifying deviations that suggest malicious activity.

Deception technology has advanced with the development of sophisticated honey pots and dynamic decoys that mimic real systems and users. These decoys not only lure attackers away from critical assets but also collect valuable intelligence on the attackers' methods and techniques, allowing for improved defensive strategies.

Security Orchestration, Automation, and Response (SOAR) platforms are transforming incident response by automating routine security tasks and orchestrating complex workflows across different security tools. This automation significantly reduces response times and enhances the efficiency of security operations, allowing human analysts to focus on more strategic tasks.

Continuous security monitoring is essential for maintaining a proactive security posture, involving real-time collection and analysis of security data to detect threats early. Compliance management is also becoming increasingly automated, with tools that ensure organizations meet regulatory requirements and industry standards, reducing the risk of non-compliance and enhancing overall security[44].

Threat simulation and red teaming provide organizations with realistic assessments of their security defenses. Threat simulations, or breach and attack simulations (BAS), continuously test security controls, while red teaming involves ethical hackers conducting simulated attacks to identify vulnerabilities. These proactive measures help organizations strengthen their defenses and prepare for real-world cyber-attacks.

## 4.3 Security Policies and Practices

The future of security policies and practices is poised for significant advancements, driven by the need to address increasingly complex and sophisticated cyber threats. One of the primary shifts will be the widespread adoption of Zero Trust Architecture, which eliminates the notion of a trusted internal network and instead requires continuous verification of all users and devices, regardless of their location. This model emphasizes stringent access controls, micro-segmentation of networks, and adaptive authentication mechanisms, thereby minimizing the risk of unauthorized access and lateral movement within the network[45], [46].

Artificial Intelligence (AI) and Machine Learning (ML) will revolutionize threat detection and response by enabling predictive analytics, automated threat hunting, and real-time anomaly detection. These technologies can analyze vast amounts of data to identify potential threats before they materialize, allowing organizations to implement preemptive security measures. Automated response capabilities, provided by Security Orchestration, Automation, and Response (SOAR) platforms, will streamline the incident response process, enabling rapid containment and remediation of threats, thus reducing the impact and duration of security breaches.

Data protection and privacy will become increasingly critical, driven by stringent regulations such as the General Data Protection Regulation (GDPR) and the California Consumer Privacy Act (CCPA). Organizations will need to develop comprehensive data governance policies that include data classification, encryption, and continuous compliance monitoring[47], [48]. Advanced encryption techniques, such as post-quantum cryptography, will be necessary to protect sensitive data against future quantum computing threats.

Continuous and role-specific employee training and awareness programs will be essential to mitigate human-related risks. These programs will incorporate behavioral monitoring tools to detect and correct risky behaviors in real-time, fostering a culture of security within the organization. Supply chain security will also gain prominence, with organizations implementing rigorous third-party risk management protocols to ensure that suppliers and partners adhere to stringent security standards. Contracts with third-party vendors will include specific security requirements and compliance mandates to protect against supply chain-related cyber threats.

Cloud security will be a major focus, with Cloud Security Posture Management (CSPM) tools becoming essential for monitoring and securing cloud environments. Implementing Zero Trust principles in multi-cloud and hybrid setups will ensure consistent security policies across diverse platforms. Blockchain technology will be leveraged to enhance security in areas such as identity management, secure transactions, and tamper-proof record-keeping, providing a robust framework for verifying and protecting digital identities and data integrity.

Moreover, the integration of emerging technologies like AI and blockchain will necessitate the development of new security policies that address ethical considerations, transparency, and accountability. Policies will need to ensure that AI and ML systems are used responsibly, with measures in place to prevent biases and ensure decision-making processes are transparent and fair. Blockchain's immutable ledger capabilities will be utilized to enhance the security and transparency of critical processes, such as supply chain management and digital identity verification[49].

The future directions of security policies and practices will be characterized by the integration of advanced technologies, adaptive security frameworks, and comprehensive regulatory compliance measures. These developments will drive the creation of dynamic, resilient security policies capable of adapting to the fast-evolving cyber threat landscape, ensuring the protection, integrity, and continuity of digital assets and operations.

# 5 Conclusion

The evolution of website information security is a testament to the ongoing battle against ever-changing cyber threats. From the foundational days of ARPANET and the development of TCP/IP to the sophisticated multi-layered security practices of today, cybersecurity has continually adapted to meet new challenges. The historical milestones, including the introduction of public-key cryptography, SSL, and the emergence of antivirus programs, have all played crucial roles in shaping modern security measures.

Today, the integration of advanced technologies such as Artificial Intelligence (AI), Machine Learning (ML), and Zero Trust Architecture highlights the dynamic nature of cybersecurity. These innovations, alongside the critical importance of cloud security and DevSecOps, illustrate the industry's proactive approach to threat detection and response. Regulatory frameworks like GDPR and CCPA enforce stringent data protection standards, driving organizations to adopt comprehensive security measures.

Looking ahead, the future of website information security promises further advancements driven by emerging technologies such as quantum computing, blockchain, and advanced encryption techniques. These innovations, coupled with enhanced international cooperation and standardization efforts, will be crucial in addressing the increasingly sophisticated cyber threat landscape.

Ultimately, the ongoing research, innovation, and collaborative efforts in cybersecurity are essential to protecting sensitive information and maintaining trust in the digital world. As cyber threats continue to evolve, so must our defenses, ensuring that we are always prepared to safeguard our digital assets and secure our interconnected lives.